\documentclass[aps,pra,twocolumn,
					showpacs,superscriptaddress,nofootinbib]{revtex4-1}

\usepackage{helvet}
\usepackage{eurosym}
\usepackage{color}
\usepackage[latin1]{inputenc}
\usepackage{amssymb}
\usepackage{amsmath}
\usepackage{graphicx}
\usepackage{bm}
\usepackage{xcolor}

\usepackage{braket}

\usepackage{ulem}

\usepackage{natbib}

\DeclareGraphicsExtensions{%
 .png,%
 .jpg,%
 .pdf,.PDF,%
 .mps,.jpeg,.jbig2,.jb2,.JPG,.JPEG,.JBIG2,.JB2}

\newcommand{\abs}[1]{\left| #1 \right|} 
\newcommand{\avg}[1]{\left< #1 \right>} 

\newcommand{\rb}[1]{\left( #1 \right)}
\newcommand{\rbb}[1]{\left[ #1 \right]}

\newcommand{\commut}[1]{\left[ #1 \right]}

\newcommand{\curlys}[1]{\left\{ #1 \right\}}

\DeclareMathAlphabet{\pazocal}{OMS}{zplm}{m}{n}
\newcommand{\super}[1]{{\pazocal{#1}}}

\def\ii{{\rm i}}

\begin{document}
\title{Wiseman-Milburn Control for the Lipkin-Meshkov-Glick Model}

\author{Sven Zimmermann}
\affiliation{Institut f\"ur Theoretische Physik,
 Technische Universit\"at Berlin,
 D-10623 Berlin,
 Germany}
\author{Wassilij Kopylov}
\email{kopylov@itp.tu-berlin.de}
\affiliation{Institut f\"ur Theoretische Physik,
 Technische Universit\"at Berlin,
 D-10623 Berlin,
 Germany}
\author{Gernot Schaller}
\affiliation{Institut f\"ur Theoretische Physik,
 Technische Universit\"at Berlin,
 D-10623 Berlin,
 Germany}
\date{\today}

\begin{abstract}
We apply a measurement-based closed-loop control scheme to the dissipative Lipkin-Meshkov-Glick model.
Specifically, we use the  Wiseman-Milburn feedback master equation to control its quantum phase transition.
For the steady state properties of the Lipkin-Meshkov-Glick system under feedback we show that 
the considered control scheme changes the critical point of the phase transition. 
Finite-size corrections blur these signatures in operator expectation values but entanglement measures such as concurrence
can be used to locate the transition point more precisely.
We find that with feedback, the position of the critical point can be shifted to smaller spin-spin interactions, which is potentially
useful for setups with limited control on these.

\end{abstract}

\maketitle


\section{Introduction}

Quantum criticalities arising in different types of phase transitions such as lasing transitions or quantum phase transitions (QPTs) belong to classical 
but still modern and important research.
This is particularly true in non-equilibrium setups~\cite{Haken-Laser_theory,Scully-quantum_optics,Phase_Transition-Theory_of_critical_phenomena-Hohenberg,Sachdev-QPT}. 
Whereas in case of the laser transition the coherence is shared by the photons that induce a macroscopic occupation of a light mode and the inverted atoms serve just as a battery, 
in case of the Hepp-Lieb quantum phase transition the collective coupling between different atoms is the key to the light flash. 

The prominent examples for collective critical models are the Dicke and Lipkin-Meshkov-Glick (LMG) models, 
where $N$ collectively coupled excited atoms in non-equilibrium produce a light flash with $N^2$ intensity during their 
decay~\cite{Dicke-Dicke_Modell,LMG-lipkin1965validity}. 
Collectivity is at the heart of such systems, and criticality is even present without any coupling to an environment already in the closed forms of such models. 
These are well studied and their transitions are known as the Hepp-Lieb QPT from a normal to a superradiant or symmetry-broken state, 
where the phases are separated by a closing energy gap between the two lowest energy states.
Additionally, the quantum fluctuations diverge at the phase transition~\cite{Hepp_Dicke-non-RWA,Hepp1973360,Narducci-spectrum_tavis_cumming,Wang-phasetransition_dicke_modell,Clive-Brandes_Chaos_and_qpt_Dicke,LMG-thermodynamical_limit-Mosseri,LMG-Finite_size_scaling-Vidal}. 

In the last decades, the fate of QPTs and their impact on the emission properties have been studied in open and non-equilibrium set-ups far from 
thermal equilibrium~\cite{Dicke_open-critical_exponent_of_noise-Nagy,Bhaseen_dynamics_of_nonequilibrium_dicke_models,Kopylov_Counting-statistics-Dicke,LMG-collective_and_independent-decay-Lee,Morrison-Dissipative_LMG-and_QPT,Ising_criticality_in_transport-Vogl}. 

Due to the high degree of parameter control in already existing experimental cold-atom realizations of such 
models~\cite{Baumann-Dicke_qpt,LMG-Exp_Bifurcation_rabi_to_jesophson-Oberthaler} new ideas have come up, particularly to study the effect of control loops on QPTs.  

In general, one can divide control in two kinds, closed-loop (feedback) control and  open control. 
Here, the difference is that closed-loop control in some way feeds information on a system state back into the system. 
In quantum (classical) mechanics, feedback control is usually further subdivided into coherent (autonomous) and measurement-based (external) feedback control. 
Many investigations have been performed by applying such schemes to quantum systems. 
For example, using an open control by periodically driving one parameter, new quantum phases could be created, entanglement modified or the emission properties changed~\cite{Extracavity_radiation_from_single_qbit-Liberato,Dicke-nonequilibrium_qpt-bastidas,Dicke-Robust_quantum_correlation_with_linear_increased_coupling-Acevedo}.
Similar and other effects like modification of the non-equilibrium steady state properties are possible with closed control loops, too.
This is also the case when an additional time-delay term is taken into account~\cite{Kopylov-time_delayed_control_Dicke,Kopylov_two-mode-TC-with-Pyragas-feedback,Feedback-two-photon-emission_droenner,Feedback-mirror_propiertes_as_time_delay_fb-carmele,Kabuss-quantum_feedback_anal_study_and_rabi_osci,Feedback_of_photon_statics-Carmele,Grimsmo-time_delayed_quantum_feedback_control}. 

Wiseman-Milburn feedback is one special example of a measurement-based control loop.
Here, the idea is to continuously monitor the reservoir coupled to the system and to perform a quantum operation on the system whenever a special event 
like a photon emission is observed~\cite{Wiseman-Quantum_measurment_control}. 
In certain limits, such control operations appear on the level of the master equation as a simple unitary modification of the jump 
term~\cite{Wiseman-Quantum_theory_of_contin_feedback}. 
Wiseman-Milburn feedback can for example be used to stabilize pure states of different systems even in 
non-equilibrium~\cite{Wiseman-Feedback_stabilization_of_pure_state,Poeltl_pure_state_stability_by_Feedback,Feedback-reverse_quant_engineering_electron_loops-Kiesslich}, 
which has been experimentally demonstrated~\cite{Feedback-Exp-Wiseman-solid_state_qubit_projective_measurment-DiCarlo,Feedback-Exp-Wiseman-control_superconducting_qubit-Huard}. 
Moreover, the impact of such schemes on the first and second laws of thermodynamics~\cite{Feedback_Thermodynam_Quant_Jump_conditioned_Strasberg} has been studied.

In this paper, our main idea is to apply Wiseman-Milburn feedback to the dissipative LMG system, where dissipation effectively corresponds to photon 
emission~\cite{Morrison-Dissipative_LMG-and_QPT,Morrison-Collective_spin_system-QPT_and_entaglement}. 
Experimentally, the scheme requires to apply after each photon emission a unitary kick along one of the three collective angular momentum axes.
Formally, this corresponds to a rotation of the jump part in the master equation around the chosen axis. 
The rotation angle is then the feedback control parameter. 
We will concentrate on the non-equilibrium steady state of the finite-size LMG system in presence of the considered 
feedback action.
In particular, we show for finite-size LMG models that in presence of feedback the non-trivial steady state expectation values of the spin operators can be reached for a smaller spin-spin coupling. 
This may be useful as an additional control knob in experiments investigating critical behaviour.
We argue that in the thermodynamic limit, the dissipative QPT, which in absence of feedback can be observed already with mean-field methods, becomes shifted toward 
a smaller spin-spin coupling as well. 
Furthermore, we show that such a shift is visible in the entanglement properties of the LMG system with feedback, too, 
which we will calculate in the dissipative context using the concurrence~\cite{Concurrence-arbitary_two_qubits_Wooters,LMG-Finite_size_scalling_Dusuel}.   

Our paper is structured as follows. 
In sections~\ref{ssec:closed} and~\ref{ssec:open} we will review the major properties of the closed and open LMG system, 
which are important to understand the feedback-induced effects. 
In section~\ref{ssec:control} we will introduce the feedback scheme. 
In section~\ref{sec:results} we will show how such a control scheme changes the steady state. 
We will investigate the properties of the spin expectation values and the concurrence. 
Finally, in section~\ref{sec:discuss} we will conclude and sum up the results.


\section{Model}
\label{sec:model}
\subsection{Closed System}
\label{ssec:closed}

The fully anisotropic Lipkin-Meshkov-Glick (LMG) Hamiltonian is given by~\cite{LMG-Critical_scaling_law_entaglement-Vidal}
\begin{equation}
\label{hamiltonian}
  H_{LMG} = - h J_z - \frac{\gamma_x}N J_x^2\,,
\end{equation}
where $\gamma_x$ is the interaction between each pair of the $N$ system spins and
$h$ is the strength of the magnetic field in $z$ direction. 
Due to the all-to-all coupling, the collective angular momentum operators can be used
\begin{align}
\label{ang_momentum_def}
  J_\eta &= \frac12 \sum_{n=1}^N \sigma_\eta^{(n)}, \quad \eta \in \curlys{x, y, z}\,,\\
  J_\pm &= J_x \pm \ii J_y\,, \notag
\end{align}
where $\sigma_\eta^{(n)}$ are the common Pauli matrices of the $n$-th spin.

The collective angular momentum operators mirror the properties of the single
spin angular momentum $[J_x, J_y] = \ii J_z$, and their action on the eigenvectors $\ket{j, m}$ is
given by
\begin{align}
\label{ang_momentum_evs}
J^2 \ket{j, m} &= j(j+1) \ket{j, m}\,,\\
  J_z \ket{j, m} &= m \ket{j, m}\,, \notag
\end{align}
where $j\in\curlys{0,\ldots,N/2}$  is the total angular momentum quantum number 
which can only assume half-integer or integer values, 
and the projection $m$ of the angular momentum to the z-axis bounds the second quantum number by $m \in \curlys{-j, j}$.
The ladder operators act as follows
\begin{equation}
\label{ang_momentum_ladder}
J_\pm \ket{j, m} = \sqrt{j(j+1) - m(m \pm 1)} \ket{j, m \pm 1}\,.
\end{equation}
We observe that the angular momentum $J^2 = J_x^2 + J_y^2 + J_z^2 = \frac{1}{2} \commut{J_-,J_+}_+ + J_z^2$ is conserved in case of the LMG system, as
it commutes with the systems Hamiltonian~\eqref{hamiltonian}. 
This follows directly from $[J^2, J_\eta] = 0$.
Therefore, the Hilbert space for different angular momenta $j$ decouples. 
Throughout this paper, we will  restrict ourselves to the part of the Hilbert space with the maximum angular momentum 
\begin{equation}
j = \frac N2\,,
\end{equation}
as this subspace contains the ground and the first excited state.
Furthermore, this subspace can be specifically realized in the experimental 
setups~\cite{LMG-Finite_size_scaling-Vidal,ESQPT-inmany_body_systems-Caprio,LMG-Exp_Bifurcation_rabi_to_jesophson-Oberthaler}. 

Another conserved quantity in the closed LMG model is the parity operator~\cite{LMG-from_perspective_of_SU11-model-Dukelsky,LMG-Nonadiabatic_dynamics_of_ESQPT-kopylov}, 
and we can define projectors onto the subspaces with positive or negative parity via
\begin{equation}
\label{parity}
  P_\pm = \frac 12 \left[\mathbf{1} \pm \exp(\ii \pi J_z) \exp(\ii \pi N/2)\right]\,.
\end{equation}

\begin{figure}[t!]
\includegraphics[width=1 \linewidth]{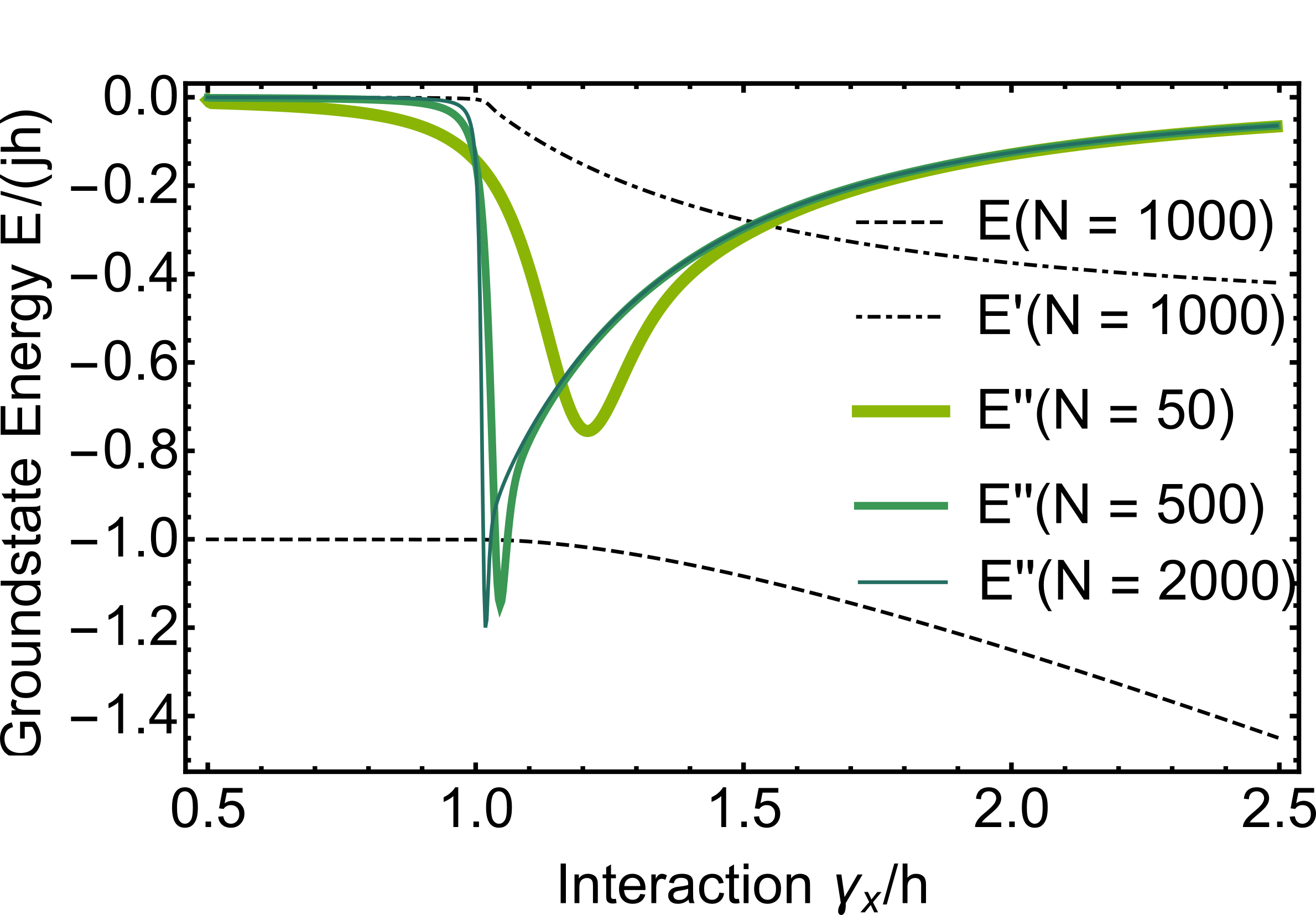}
\caption{Plot of the LMG ground state energy density and its first two derivatives for different $N$. 
For a larger $N$, the minimum of the second derivative if further shifted to $\gamma_x/h = 1$ and becomes sharper, see green solid lines with decreasing thickness.
In the thermodynamic limit $N \to \infty$, the second derivative is not continuous any more and thus the ground state is not 
analytic at $\gamma_x = h$ which indicates that the phase transition is of second order.
}
\label{fig:ground_state_energy}
\end{figure}

One of the intriguing properties of the LMG model is the appearance of a quantum phase transition (QPT) 
in the thermodynamic limit $N \to \infty$ at a critical value of the coupling 
$\gamma_x/h = 1$~\cite{LMG-phase_transition-Gilmore,LMG-grouns_state_studies_coherent_states-Gilmore,LMG-Finite_size_scaling-Vidal,LMG-spectrum_thermodynamic_limit_and_finite_size-corr-Mosseri}. 
Fig.~\ref{fig:ground_state_energy} shows the ground state energy per atom pair (dashed line) and its first two derivatives. 
Also the second derivative  $E{''}$ with respect to $\gamma_x$ is shown for different spin numbers $N$ (solid green curves with different thickness), and it becomes discontinuous in the thermodynamic limit 
at $\gamma_x/h = 1$, which indicates the appearance of the second order QPT in the ground state energy per atom pair. 

\begin{figure}[t!]
\includegraphics[width=1 \linewidth]{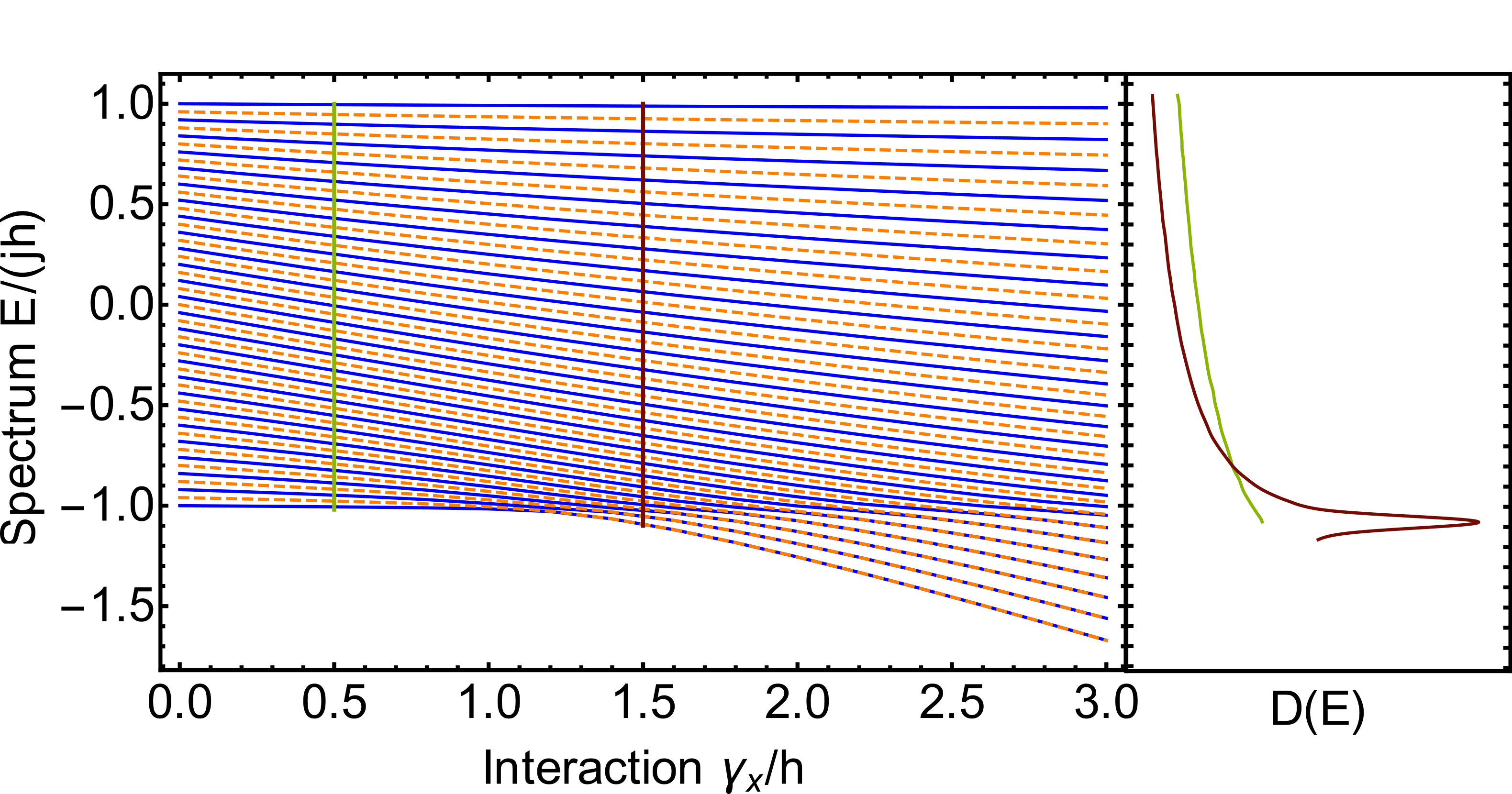}
\caption{(left) Energy spectrum of the LMG Hamiltonian for $N = 50$ versus the interaction $\gamma_x$.
  The colors indicate the parity of the state: blue (solid) for positive and orange (dashed) for negative parity.
  The QPT precursor is visible around $\gamma_x \approx 1.3 h$ due to finite size effects. 
  For $N \to \infty$ this will shift to $\gamma_x = h$ (compare with Fig.~\ref{fig:ground_state_energy}).
  Additionally, the higher density of states around $E = -jh$ signals the ESQPT.
  (right) The density of states versus the energy at $\gamma_x = 0.5 h$ (green) and
  $\gamma_x = 1.5 h$ (red) for $N = 1000$ as indicated in the left panel.
  The peak is visible only for $\gamma_x > h$ and corresponds to the ESQPT.}
\label{fig:spectrum_closed}
\end{figure}

However, the excited part of the LMG spectrum has an additional phase transition -- the excited states quantum phase transition (ESQPT)~\cite{ESQPT-inmany_body_systems-Caprio,ESQPT_Docoherence_two_level_boson_pedro,brandes2013excited,ESQPT-Structure_eigenstate_and_quench_dynamic-Santos,ESQPT-system_with_two_freedom_degrees_finite_size-Cejnar,LMG-TC-periodic_dynamic_and_QPT-Georg}. 
It is visible in Fig.~\ref{fig:spectrum_closed}, which shows the spectrum of the LMG model for $j=N/2$ as a function of $\gamma_x$ for $N = 50$. 
The signature of the ESQPT is the non-analyticity in the density of states
\begin{equation}
\label{density_of_states}
  D(E) = \frac1V \sum_{i = 1}^{N+1} \delta(E - \epsilon_i)\,,
\end{equation}
which is visible in the symmetry-broken phase. 
Here, $V$ is a normalization constant and $\epsilon_i$ is the $i$th eigen-energy of the system. 
Although strictly speaking the non-analyticity appears only in the thermodynamic limit, in case of the LMG model it is already visible 
in Fig.~\ref{fig:spectrum_closed}(left) as a dense region in the spectrum around $E/(jh) = -1$. 
The right panel of Fig.~\ref{fig:spectrum_closed} shows the density of sates numerically determined for $N = 1000$ along the green and brown 
vertical lines indicated in the left panel. 
For $\gamma_x/h > 1$ (brown curve) the density of states has a logarithmic divergence, which marks the
ESQPT~\cite{ESQPT-inmany_body_systems-Caprio,ESQPT_imparct_of_qpt_on_level_dynamics-Cejnar,QPT_interacting_boson_model-Cejnar,LMG-Critical_scaling_law_entaglement-Vidal,brandes2013excited}.

\subsection{Open System}
\label{ssec:open}

The dynamics of an open LMG system can be described in the master equation formalism according to Ref.~\cite{Morrison-Dissipative_LMG-and_QPT} via
\begin{equation}
 \label{eq:master_equation}
 \dot\rho = -\ii \commut{H_{\rm LMG}, \rho} + \frac{\kappa}{N}  \super{D} [J_+] \rho\,.
\end{equation}
Here, $\kappa\ge 0$ quantifies the strength of dissipation, which as a super-operator acts as
\begin{equation}
\label{eq:master_equation1}
    \super{D}[J_+] \rho = \frac12 \left(2 J_+ \rho J_- - J_- J_+ \rho - \rho J_- J_+\right)\,.
\end{equation}
For $h>0$ and $\gamma_x=0$ we see that this dissipator has the pure state $\ket{j,+j}$ as a stationary state, which is the ground state of $H_{\rm LMG}$ for $\gamma_x = 0$ and $h>0$. In general, the steady state will be a mixed non-equilibrium state. 

The master equation can be rewritten in terms of the super operators~\cite{Breuer-open_quantum_systems,Schaller-QS_far_from_equilibrium} 
\begin{equation}
\label{master_equation_supOp}
\dot \rho = \super{L} \rho = \super{L}_0 \rho + \super{J} \rho,  
\end{equation}
with the free Liouvillian $\super{L}_0 \rho = -\ii\rb{H_{eff} \rho - \rho H_{eff}^\dagger}$, which describes the system evolution without jumps, 
and the jump superoperator $\super{J} \rho  = \frac \kappa N J_+ \rho J_-$. 
Note that here, $H_{eff}$ is an effective non-hermitian Hamiltonian~\cite{Breuer-open_quantum_systems,Jung-Phase_transition_open_qs,Kopylov-LMG_ESQPT_control}
\begin{equation}
\label{eq:H_nonhermit}
H_{eff} = H_{LMG} - \frac{\ii}{2} \frac\kappa N J_- J_+\,.
\end{equation}
The total angular momentum $J$ is conserved even in presence of dissipation ($\kappa > 0$), but not the parity. 
Due to the presence of the all-to-all coupling, the dynamics of the observables governed by the master equation~\eqref{eq:master_equation} 
can be very well described using only the first-order moments in mean-field approximation 
$\avg{J_\eta \cdot J_\eta'} \approx \avg{J_\eta} \cdot \avg{J_\eta'}$~\cite{LMG-Finite_size_scalling_Dusuel,Morrison-Collective_spin_system-QPT_and_entaglement,LMG-TC-periodic_dynamic_and_QPT-Georg}.  
The approximation yields up to four different steady-state solutions. 

\begin{figure}[ht]
\includegraphics[width=1 \linewidth]{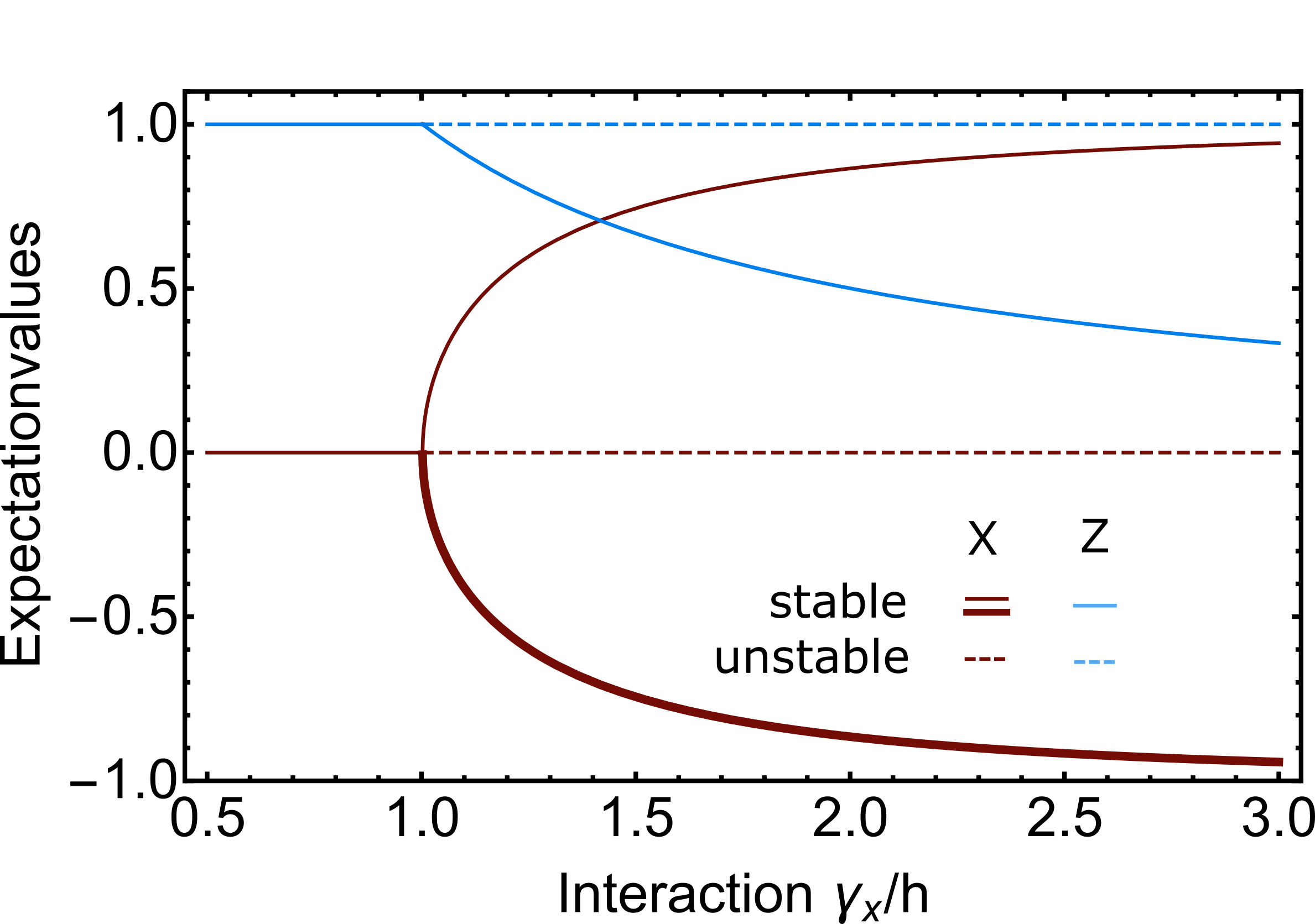}
\caption{\label{fig:mean_field}
Plot of the mean-field stationary large-spin expectation values $X = \avg{J_x}_{t\to \infty}/j, Z = \avg{J_z}_{t \to \infty}/j$ versus spin-spin interaction strength $\gamma_x$ in the thermodynamic limit for $\kappa/h = 0.05$. The stable trivial solution with $Z = 1$ (solid) becomes unstable at $\gamma_x^{\rm cr}$ from Eq. \eqref{eq:qpt_position} (dashed). But two new stable solutions  (distinguishable by $X$)  are created by a pitchfork bifurcation for $\gamma_x > \gamma_x^{\rm cr}$ (solid) with $Z < 1$. 
}
\end{figure}

In presence of dissipation, the critical point is now defined by the stability exchange of those steady states at
\begin{equation}
\label{eq:qpt_position}
  \gamma_x^{\rm cr} = h + \frac{\kappa^2}{4 h}\,.
\end{equation}
Fig.~\ref{fig:mean_field} summarizes the steady state properties governed by the mean-field equations. 
The QPT is in this picture represented by a bifurcation, visible in the spin expectation values. 
While in the normal phase for $\gamma_x < \gamma_x^{\rm cr}$ there is one stable (solid) and one unstable (not shown) fixed point, in the symmetry broken phase for $\gamma_x > \gamma_x^{\rm cr}$ 
the stable fixed point of the normal phase becomes unstable (dashed lines), and two new stable fixed points appear (solid curves with different thickness).
Thus at $\gamma^{\rm cr}$ the solution properties change drastically, the stable solution (solid) of the normal phase becomes unstable (dashed) 
and a new stable solution is created via a pitchfork bifurcation (solid). More details are presented in App.~\ref{appendix}.  

\subsection{Controlled System}
\label{ssec:control}

In this section, we extend the dynamics of the open LMG system by the Wiseman-Milburn control operation~\cite{Wiseman-Quantum_measurment_control}. 
The idea is to apply an instantaneous unitary control operation $\super{C}$ on the density matrix after each jump $\super{J}$
\begin{equation} 
\label{eq:control_c}
\super{C} \rho = U_C \rho U_C^\dagger\,.
\end{equation}

\begin{figure*}[ht]
	\includegraphics[width=0.48 \linewidth]{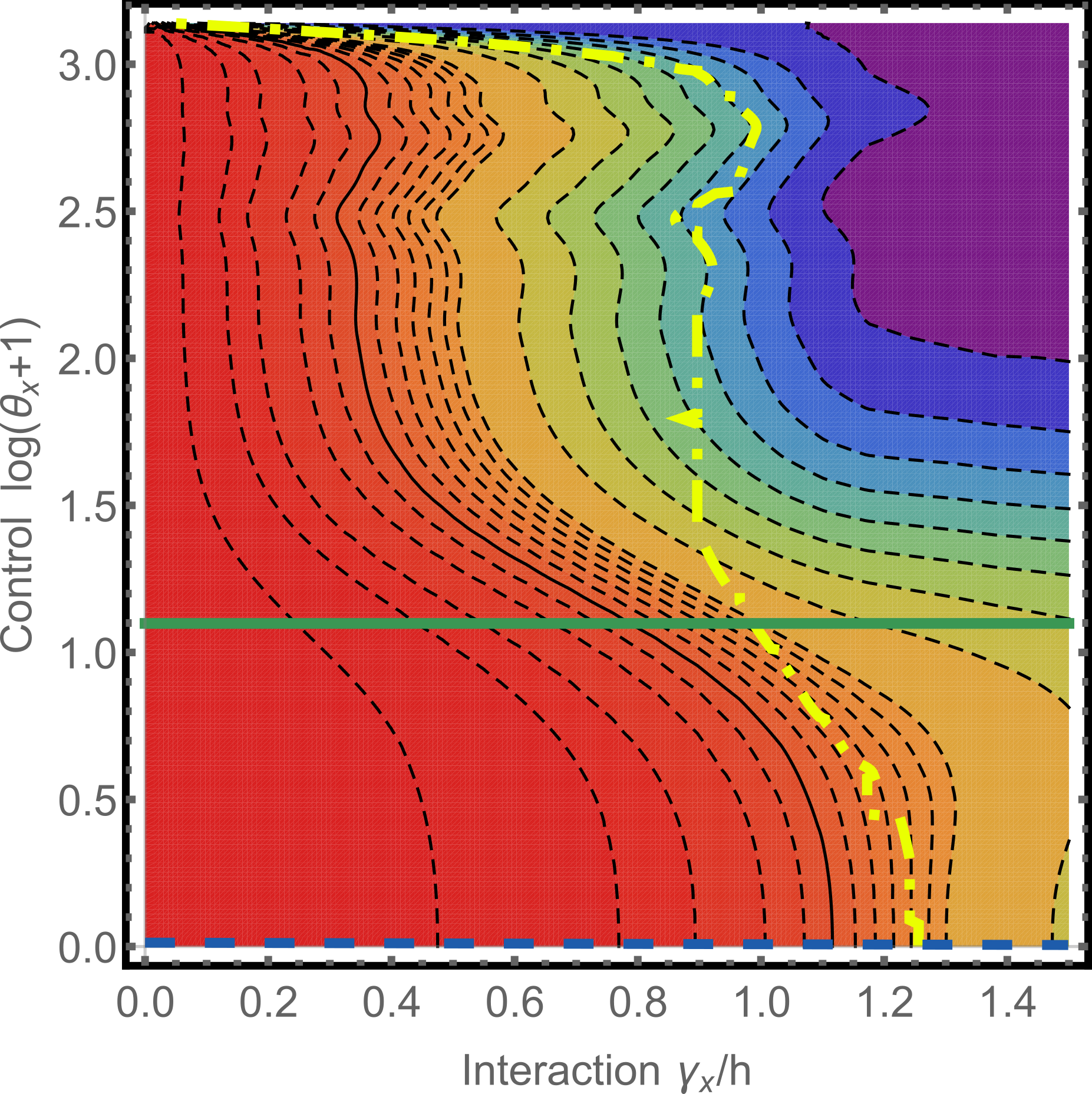}
	\includegraphics[width=0.49 \linewidth]{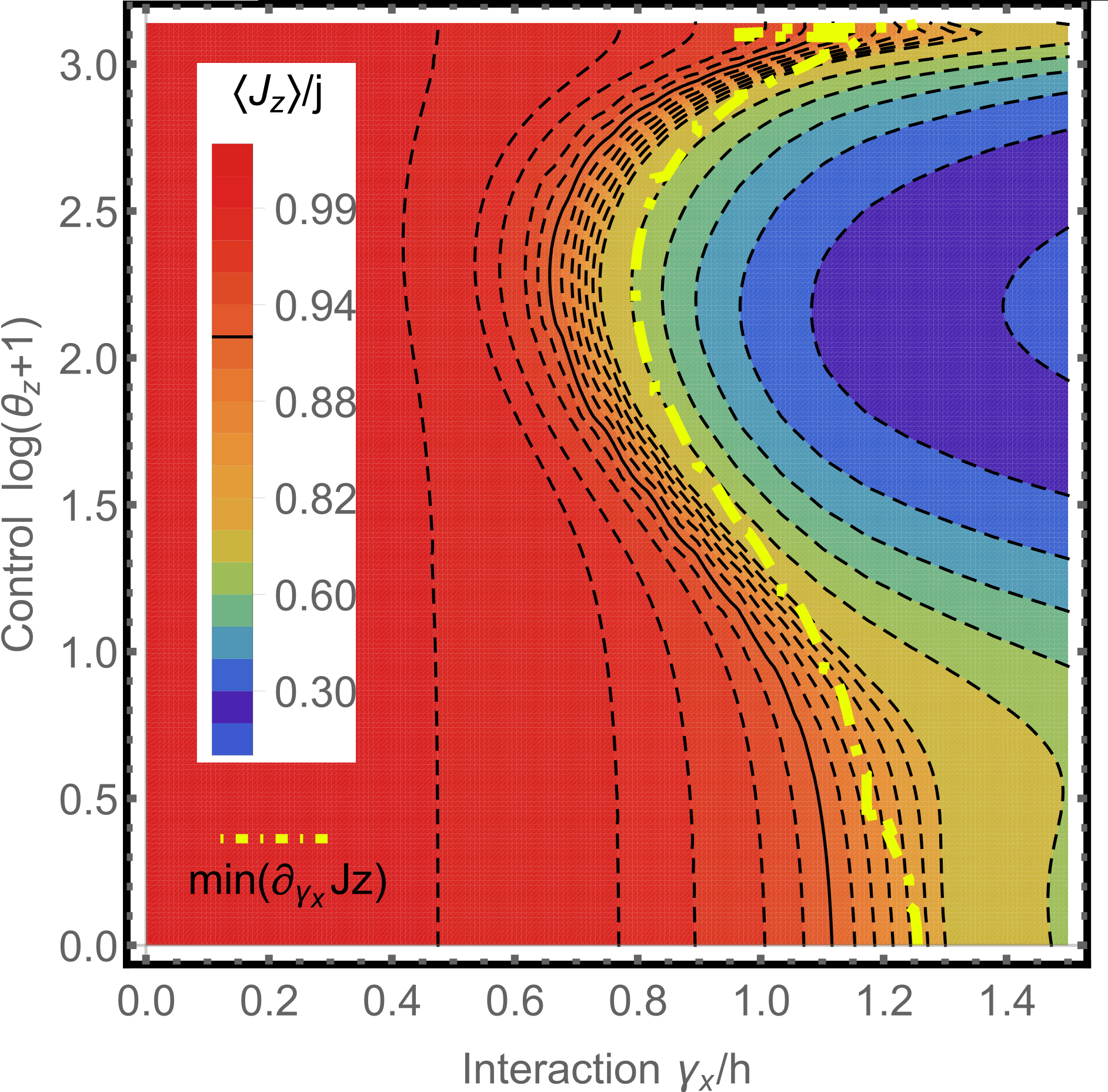}
	\caption{The normalized expectation values $\avg{J_z}/j$ versus the interaction $\gamma_x$ and the control angle $\theta_x$ (left) or $\theta_z$ (right)
		for $N = 50$ and $\kappa = 0.05 h$. 
		The solid contour line shows the $\avg{J_z} = 0.93$-value where the finite-size QPT occurs in the $\theta = 0$ case.
		The dot-dashed line shows the position of the minimum in the derivative of $\avg{J_z}$ with respect to $\gamma_x$. 
		Increasing the control parameter $\theta$ shifts the onset of the transition towards the smaller $\gamma_x$ values, see the trend of the 
		contour lines around the solid one. In case of the $J_z$ control (right) this trend is reversed for higher $\theta_z$ values. 
		However, the critical point can be determined only approximately due to the presence of finite-size effects. 
		The $\avg{J_z}$ values and their derivatives along the horizontal dotted $(\theta_x = 0)$ and solid lines $(\theta_x = 2)$ in the left contour plot are shown in Fig.~\ref{fig:jz_expectation_contour}. Note the logarithmic scaling of the y-axis, thus the maximal $\theta_\eta$ value which we use is $\pi \sqrt{N} \approx 22$. 
		}
	\label{fig:jz_expectation_contour}
\end{figure*}

Such a measurement-based feedback can be described on the master equation level, where Eq.~\eqref{master_equation_supOp} is altered to~\cite{Wiseman-Quantum_measurment_control,Poeltl_pure_state_stability_by_Feedback,Feedback_Thermodynam_Quant_Jump_conditioned_Strasberg},
\begin{equation}
\label{eq:liouville_control}
 \dot\rho =  \super{L}\rho = \super{L}_0 \rho + \super{C} \super{J} \rho\,.
\end{equation}
We consider simple rotations conserving the total angular momentum
\begin{equation}
\label{eq:control_uc}
  U_C = \exp\rbb{-\frac{\ii}{\sqrt{N}} \rb{\theta_x J_x + \theta_y J_y + \theta_z J_z}}\,,
\end{equation}
where the $\theta_\alpha$ are real control parameters and where we have introduced the factor $\sqrt{N}$ to ensure for
convergence of stationary expectation values in the thermodynamic limit.

In the following, we will always use either $\theta_x \neq 0$ or $\theta_z \neq 0$, as other combinations are either producing similar effects as rotation around $x$ or $z$ axis 
or could not improve them. 
Usually, the  Wiseman-Milburn scheme can be used to stabilize the eigenstates of the effective non-hermitian Hamiltonian Eq. \eqref{eq:H_nonhermit}~\cite{Feedback-reverse_quant_engineering_electron_loops-Kiesslich,Poeltl_pure_state_stability_by_Feedback} by a rotation of the state after each jump to such an eigenstate. 
However, in our case this is not applicable, as higher-order terms like $J_\eta^n J_{\eta'}^{n'}$ would have to be used to achieve this. 
Despite this restriction, we will show in the next section that even a single $J_x$ or $J_z$ rotation can dramatically change the systems steady state,  
modify its entanglement and shift the point of the phase transition. 

Before proceeding to the results, we note that a simple mean-field analysis is not applicable here.
We found that in presence of feedback, the simple mean-field approximation violates the conservation of total angular momentum. Likewise, analysis of $H_{\rm eff}$ \cite{Kopylov-LMG_ESQPT_control} cannot reveal any feedback-induced effect as $H_{\rm eff}$ is insensitive to it. 
Instead, the full feedback master equation Eq.~\eqref{eq:liouville_control} has to be solved numerically for its steady state, which is more demanding and restricts our study to the finite-size regime.


\section{Impact on the Steady State}
\label{sec:results}

In this section, we discuss the steady state properties of the LMG system under the jump-based feedback action for a finite number of atoms $N$. 
In practice, we numerically determine the stationary state of Eq.~\eqref{eq:liouville_control}. 
Below, we first show how the typical observables like the spin expectation values are changed under the feedback and discuss their connection to the QPT. 
Later, we will investigate the systems entanglement.


\subsection{Observables in the new steady state}
\label{ssec:expectation}
We start our discussion by showing in Fig.~\ref{fig:jz_expectation_contour} the $\avg{J_z}$ expectation values in the steady state of Eq.~\eqref{eq:liouville_control}  as a function of the interaction strength 
$\gamma_x$ and the control angle $\theta_x$ (left, for $\theta_y=\theta_z=0$) or $\theta_z$ (right, for $\theta_x=\theta_y=0$). Note the logarithmic scaling of the y-axis.  

Our numerics shows that in presence of the considered feedback scheme for $\theta_x >0$, the decay of the $\avg{J_z}$ expectation value is observed for smaller $\gamma_x$, see Fig.~\ref{fig:jz_expectation_contour}(left). 
The contour lines are shifted then toward the smaller $\gamma_x$ values. 
For small $\theta_x$ values the shift is approximately linear, then for $1.5 < \log(\theta_x +1) < 3$ the position of $\avg{J_z}$ values does not change much. 
For even larger $\theta_x$ values the deviation of $\avg{J_z}$ from its maximal value $j = N/2$ is again strongly shifted to the left. 
For $\log(\theta_x +1) \to \log(\pi\sqrt{N} +1) \approx 3.24$ this point shifts to $\gamma_x \to 0$. 

\begin{figure}
	\includegraphics[width=0.99 \linewidth]{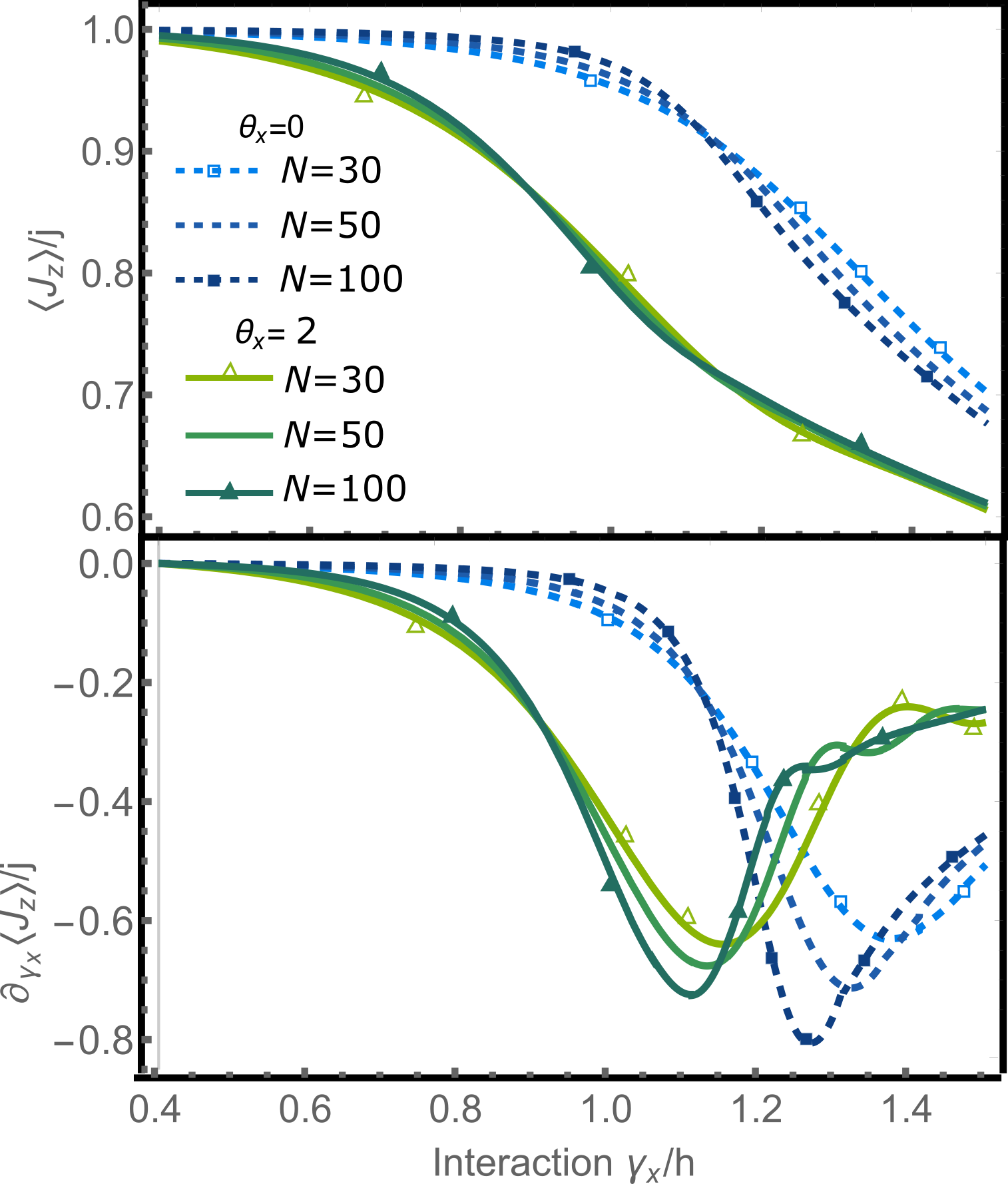}
	\caption{The expectation values $\avg{J_z}$ (top) and the derivative $\partial_{\gamma_x} \avg{J_z}$ (bottom) over the interaction $\gamma_x$ for $\theta_x = 0$ (blue dashed lines) and $\theta_x = 2$ (green solid lines) along the the horizontal lines in Fig. \ref{fig:jz_expectation_contour}(left) for N = 50. Curves marked by open (closed) symbols additionally show the finite size effects for $N = 30(100)$. Increase of the control angle $\theta_x$ shifts the point where the $\avg{J_z}$ value starts to strongly deviate from its maximum value and shifts the position of the minimum in the derivative.}
	\label{fig:jz_expectation_lines}
\end{figure}

We will elaborate this findings by comparing the  behaviour for two fixed $\theta_x$ parameters. The $\avg{J_z}(\gamma_x)$ expectation values along the dashed $(\theta_x = 0)$ and solid $(\theta_x = 2)$ horizontal lines in Fig. \ref{fig:jz_expectation_contour}(left) are shown in Fig. \ref{fig:jz_expectation_lines}(top), see dashed and solid line for $N = 50$, respectively.  In absence of feedback (dashed lines), the case is well studied~\cite{LMG-Finite_size_scaling-Vidal,LMG-dynamical_properties_accros_qpt-Mosseri} and the expectation values may serve as a signature of the QPT. 
Indeed, for an increasing number of particles $N$ the $J_z(\gamma_x)$ curves become more irregular around the critical point, 
see dotted blue lines marked by different symbols for three different values of $N$. 
The visibility of this signature is improved in the first derivative $\partial_{\gamma_x} J_z$ as a minimum, see Fig. \ref{fig:jz_expectation_lines}(bottom),
which becomes sharper with increasing $N$ and finally becomes discontinuous in the thermodynamic limit $N \to \infty$ 
at $\gamma_x = \gamma_x^{\rm cr} = h+\kappa^2/(2h)$ where the dissipative QPT occurs 
(this case is not shown in Fig.~\ref{fig:jz_expectation_lines}, see Fig.~\ref{fig:mean_field} instead).

Associating the critical point of the dissipative QPT with the $\avg{J_z}$ deviation from its maximum value means that the considered control scheme 
pushes the point of the dissipative QPT to a smaller critical value than in the $\theta = 0$ case. 
For $\theta_x = 2$ we demonstrate this behaviour by plotting $\avg{J_z}$ across the horizontal solid line in the contour plot, 
see the unmarked solid curve for $N = 50$ in Fig.~\ref{fig:jz_expectation_lines} (top). 
The expectation value $\avg{J_z}$ remains first constant with the increasing coupling $\gamma_x$, then it starts to decrease. 
The value of $\gamma_x$, where the decrease takes place, is changed by the feedback in comparison to the $\theta_x = 0$ case (dotted lines). 
The latter is especially visible in the derivative with respect to $\gamma_x$, see Fig.~\ref{fig:jz_expectation_lines}(bottom). 
That means, that the feedback forces the system to leave the usually stable solution -- after passing the now $\theta_x$-dependent critical $\gamma_x$ value --  
with its maximal $\avg{J_z}$ value and to converge to another one, with $\avg{J_z} <0$. 
The latter solution is typical for the symmetry broken phase of the uncontrolled LMG model. 
Additionally, for each curve, we show the finite-size scaling of the results by labelling the correspondent curve with open (filled) symbols for $N = 30 (100)$. 
Besides the feedback-induced shift of the dissipative QPT which was discussed above, the calculation shows that larger $N$ lead to a more discontinuous shape in the vicinity
of the critical point as expected. 
 
Due to finite size effects, the position of the shifted dissipative phase transition can only be determined approximately. 
The solid contour line in Fig.~\ref{fig:jz_expectation_contour} shows the $J_z$ value, at which the finite-size precursor of the QPT in absence of feedback takes place.
In contrast, the thick dot-dashed line in Fig.~\ref{fig:jz_expectation_contour} shows the position of the minimum of $\partial_{\gamma_x}J_z$ for different $\theta_x$ values. 
Obviously, these curves do not coincide.
Changing the number of particles $N$, see Fig.~\ref{fig:jz_expectation_lines}(bottom), or analysing e.g. the $\avg{J_x^2}$ value instead would lead to a considerable shift of the solid and dot-dashed lines in the phase diagram. 
The presence of the QPT in the Hamiltonian part leads to oscillations right from the minimum of the $\avg{J_z}$ derivative, see green lines in Fig.~\ref{fig:jz_expectation_lines} (bottom). 
We checked that the beginning of the oscillations coincides with the minima of the $\partial_{\gamma_x} \avg{J_x}$ in case without feedback. 
As the calculation for much larger $N$ becomes rather involved, we show in the next chapter that the use of concurrence allows to fix the position of 
the QPT for in presence of feedback much more precise. 
 
In case of the active $\theta_z$ control, the feedback impact on the steady state properties of $\avg{J_z}$ is qualitatively similar, but the shift is much smaller. 
Both phases are always clearly separated and the smallest $\gamma_x$ where the $\avg{J_z}$ starts to deviate from $j$ is reached around $0.8 h$ for 
$\theta_z = \frac{\pi}{2} \sqrt{N}$, see right part of Fig.~\ref{fig:jz_expectation_contour}.


\subsection{Entanglement}
\label{ssec:entanglement}

\begin{figure*}
	\includegraphics[width=0.49 \linewidth]{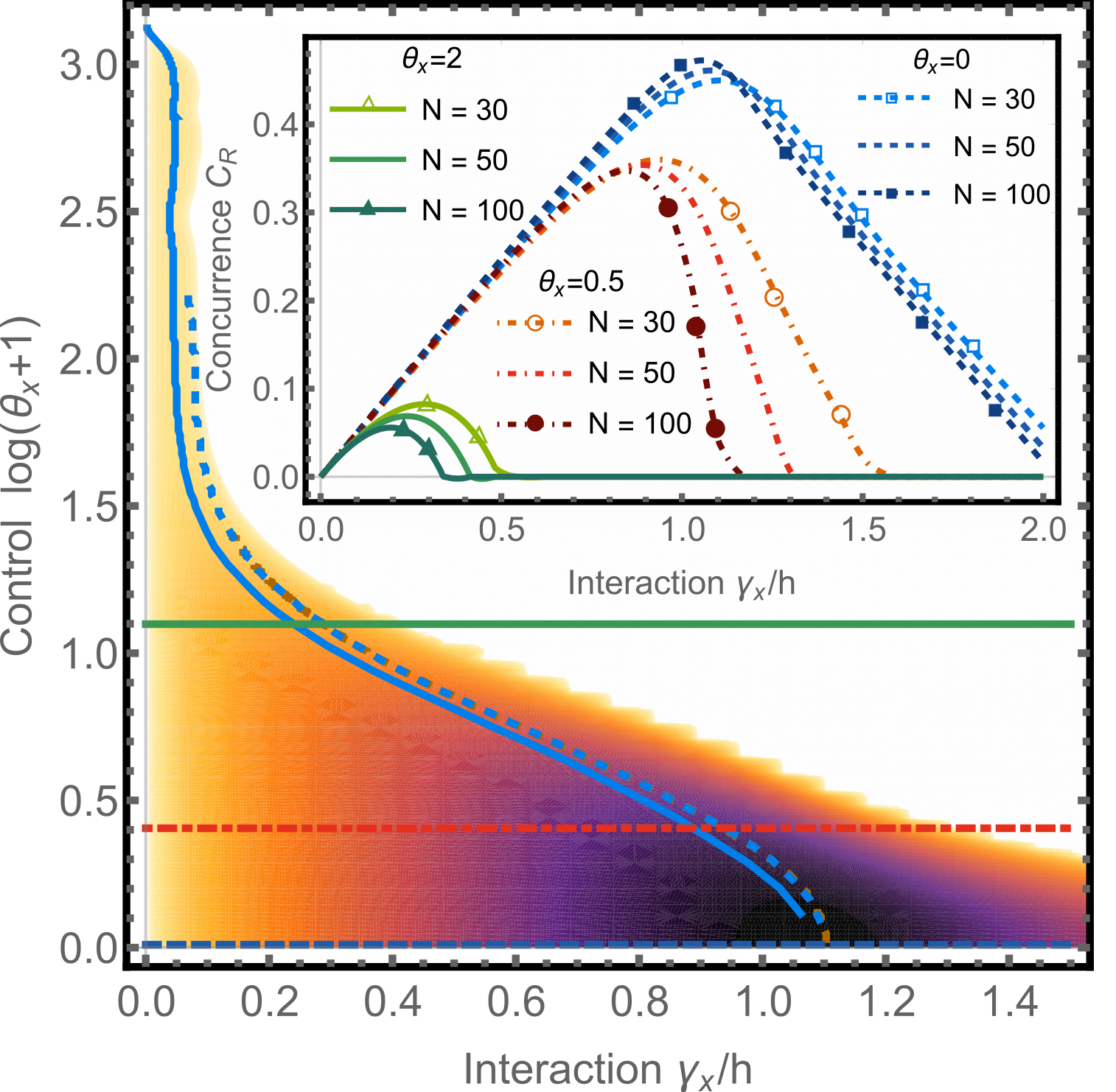}
	\includegraphics[width=0.49 \linewidth]{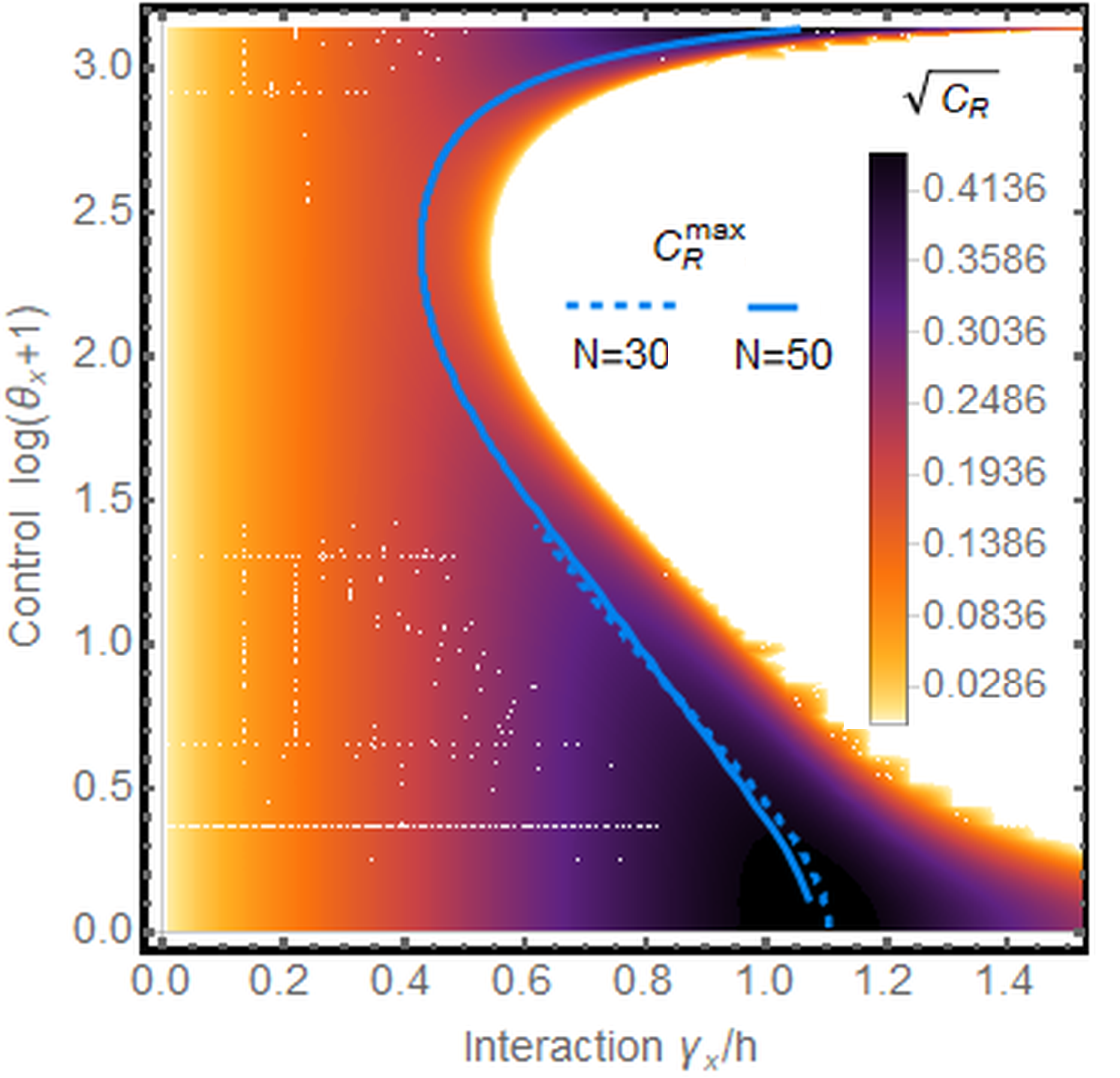}
	\caption{Plot of the concurrence over the interaction $\gamma_x$ and the control strength (note the logarithmic scaling) $\theta_x$ (left) or $\theta_z$ (right) for $N = 50$ and $\kappa = 0.05 h$. 	The blue solid line marks the region with maximum concurrence $C_R^{\rm max}$ and separates two phases of the LMG model in presence of feedback. 
	The point with the maximal concurrence is shifted to smaller $\gamma_x$ values and concurrence width shrinks with increasing control strength $\theta_x$. In contrast, in presence of the $J_z$ control (right) this trend is reversed for higher $\theta_z$ values. 
	(Inset) The inset shows the concurrence shift for different $\theta_x$ values along the blue (dashed)~\cite{LMG-Finite_size_scalling_Dusuel}, 
	red (dot-dashed) and green (solid) horizontal cuts in the left contour plot. 
	On top, we show the finite-size behaviour of the concurrence for different number of $N$, see lines marked with open $(N = 30)$ and filled $(N =100)$ symbols.
	}
	\label{fig:concurrence_contour}
\end{figure*}

Due to the presence of dissipation in our model, the stationary density matrix is not a pure state, and 
the entanglement cannot \cite{Concurrence_mixted_state-Wooters} be characterized by the common 'von Neumann entropy' 
$S(\sigma)=-{\rm Tr}\sigma \ln \sigma$ of a reduced part of the system density matrix $\sigma = {\rm Tr}_{\rm part} \{\rho\}$ \cite{LMG_Entanglement_entropy_by_tracing_out-Vidal}. 
Instead, the Wootters concurrence has to be used, which in general is very hard to calculate~\cite{Concurrence-pair_of_Qbits-wooters}. 
However, it is possible to calculate it from the reduced density matrix of two Qbits in an arbitrary basis~\cite{Concurrence-arbitary_two_qubits_Wooters}. 
The concurrence is bounded between zero and one and gives the amount of entanglement between two Qbits. 
In case of symmetric states --  to which we constrain ourselves by considering the $j=N/2$ subspace only --  it is possible to express the concurrence $C_R$ using the 
large-spin observables~\cite{Concurrence-Symmetric_states-Molmer,Morrison-Collective_spin_system-QPT_and_entaglement}.

\begin{align}
\label{eq:concurrence_def}
C_R &= 
\begin{cases}
2 \max\rbb{0,\frac{\abs{\avg{J_+^2}}}{N} - \frac{\avg{J_x^2} + \avg{J_y^2}}{N} + \frac{1}{2} }  {\rm ,}  \frac{N^2 - 4 \avg{J_z^2}}{2 N} < B,\\
2 \max\rbb{0,\frac{N}{4} - \frac{\avg{J_z^2}}{N}- S}  {\rm ,}    \frac{N^2 - 4 \avg{J_z^2}}{N} \geq B,
\end{cases} \\
S &= \frac{\sqrt{\rbb{N(N-2) + 4 \avg{J_z^2}}^2 - \rbb{4(N-1) \avg{J_z}}^2}}{4N},\notag \\
B &=  S + \frac{\abs{\avg{J_+^2}}}{N}. \notag 
\end{align}

Concurrence has been widely studied in collective systems like LMG or Dicke models with and without dissipation~\cite{Morrison-Collective_spin_system-QPT_and_entaglement,LMG-Finite_size_scalling_Dusuel,Dicke-Robust_quantum_correlation_with_linear_increased_coupling-Acevedo,Dicke_concurrence-collective-angular-momentum-Milburn,LMG_Entanglement_dynamics_Vidal,LMG_Entanglement_entropy_by_tracing_out-Vidal,Dicke_Entanglement_Lambert}. 
Here we will use the modified steady state in presence of Wiseman-Milburn feedback term to study the modification of the concurrence and compare it with the known cases.   

The blue dashed lines in the inset of Fig.~\ref{fig:concurrence_contour}(left) show the concurrence of the open system $\kappa \neq 0$ without feedback $\theta_x = 0$ 
for three different $N$-values~\cite{LMG-Finite_size_scalling_Dusuel,Morrison-Collective_spin_system-QPT_and_entaglement}. 
Increasing $\gamma_x$, the concurrence $C_R$ grows in the normal phase, reaches a maximum at the quantum-critical point, which is shifted due to the 
finite-size scaling to the right of $\gamma_x^{\rm cr}$~\cite{LMG-Finite_size_scalling_Dusuel,Morrison-Collective_spin_system-QPT_and_entaglement}.
The concurrence decays then to zero for higher $\gamma_x$ values. 
As its peak indicates the quantum-critical transition \cite{Entanglement-in_QPT_and_scaling_XXZ_chain-Gu,Entanglement-and_QPT_in_spinModels-Gu,Dicke_Entanglement_Lambert,Entanglement_second_order_QPT-Vidal,LMG-Finite_size_scalling_Dusuel}, we will use concurrence in presence of feedback for
identifying the maximum of concurrence $C_R^{\rm max}$ with a precursor of the QPT in the thermodynamic limit. 
Indeed, the maximum of the concurrence for finite $\theta_x$ values is shifted with increasing $\theta_x$ values to the smaller $\gamma_x$, 
see red (dot-dashed) and green (solid) curves in the inset without additional markers. 
The curves marked with open (closed) symbols show the finite size effects for $N = 30 (100)$. 
The critical point becomes then shifted by the feedback, too, as we have already observed in the previous section. 
The maximal concurrence $C_R^{\rm max}$ is shown in Fig.~\ref{fig:concurrence_contour}(left) along the blue solid line in Fig.~\ref{fig:jz_expectation_contour} in the $(\gamma_x,\theta_x)$ plane.
The blue dotted line shows the maximum concurrence for $N = 30$ particles. 
The finite-size shift is rather small, this suggests that the maximum of the concurrence is a more precise method to determine the final point of the dissipative QPT. 
The blue line can also be seen as the phase separation, left from the line the system is in the normal phase, right from it the system would be in the symmetry broken phase in the thermodynamic limit.
Increasing the control parameter $\theta$ additionally shrinks the zone of the non-vanishing concurrence (colored zone in Fig.~\ref{fig:concurrence_contour}). 
Our numerics show that for $\theta \to \pi\sqrt{N}$ the zone with the normal phase vanishes.
In this regime, the feedback action rotates the normal steady state immediately into a non-trivial one.
       
For completeness, the right part of Fig.~\ref{fig:concurrence_contour} shows the concurrence for the $J_z$-control. 
The form of the contour plot mimics the behaviour of the $\avg{J_z}$ values from Fig.~\ref{fig:jz_expectation_contour}(right). 
With increasing $\theta_z$, the area with non-vanishing concurrence shrinks and therefore the point of the dissipative QPT moves to the left. 
But for even larger values of $\theta_z$, this trend is reversed and the QPT moves to the right again. 
At $\theta = \sqrt{N} \pi$ the old QPT without control is recovered as can be understood analytically from the spectrum of the $J_z$ operator. Also here we observe that the maximum of concurrence is less sensitive to finite-size effects. 


\section{Summary}
\label{sec:discuss}

We have applied the Wiseman-Milburn control scheme to the dissipative Lipkin-Meshkov-Glick model by modifying the jump part of the corresponding master equation. 
The scheme monitors the reservoir and applies a kick in form of a unitary rotation around some spin axis after emission events. 
We determined numerically the steady state of the feedback master equation in the finite-size regime and used it to calculate the 
spin expectation values and the concurrence, which quantifies the spin-spin entanglement within the system.
Without feedback, the $\avg{J_z}$ expectation value stays constant (up to finite-size corrections) while increasing the spin coupling $\gamma_x$, until the 
critical point $\gamma_x^{\rm cr}$ is reached, where the concurrence reaches there its maximum value. 
For $\gamma_x > \gamma_x^{\rm cr}$ both observables start to decay. 
Our approach could reproduce these finite-size signatures of a dissipative quantum phase transition~\cite{LMG-Finite_size_scaling-Vidal,LMG-Finite_size_scalling_Dusuel,Morrison-Collective_spin_system-QPT_and_entaglement}. 
In presence of feedback, such signatures are shifted to smaller $\gamma_x$ values and the shift can be controlled by the feedback parameter, 
which is a rotation angle around one of the three spin axis. 
The concurrence values becomes smaller and the region with non-zero concurrence shrinks by an active feedback loop. 
However, as is clear from its definition, the concurrence represents only a lower bound of entanglement which one could have in a system~\cite{Concurrence-arbitary_two_qubits_Wooters}. 
The applied feedback rotation cannot rotate the system into an eigenstate of the effective Hamiltonian since these would require significantly more sophisticated control operations. 
To implement general control operations would be highly non-trivial especially for larger $N$ values.
Therefore, we could detect the signatures of the intrinsic QPT in the spin expectation values as well.
We believe that our methods can be useful in the study of other spin systems and feedback problems with delay as well.


\section*{Acknowledgements}\vspace{-2mm}
We thank Javier Cerrillo for useful discussions. The authors gratefully acknowledge financial support from the DFG (grants BR 1528/9-1, SFB 910, GRK 1558). 


\appendix*
 
\section*{Appendix}
\label{appendix}

Using $\partial_t \avg{J_\eta} = {\rm Tr}\left(J_\eta \dot \rho\right)$  
with the mean-field assumptions, the equations of motions without feedback become~\cite{Morrison-Collective_spin_system-QPT_and_entaglement,Morrison-Dissipative_LMG-and_QPT}
\begin{align} 
\label{eq:mean_field}
X &= \frac1j \braket{J_x}, Y = \frac1j \braket{J_y}, Z = \frac1j \braket{J_z} \quad\,, \\
\dot X &= h Y - \frac\kappa2 Z X \,, \notag\\
\dot Y &= -h X + \gamma_x Z X - \frac\kappa2 Z Y \,, \notag\\
\dot Z &= -\gamma_x X Y + \frac12 \rb{X^2 + Y^2} \,. \notag
\end{align}	
The equations from \eqref{eq:mean_field} have two steady-state solutions.
The first one $(X, Y, Z) = (0, 0, 1)$
is stable in the normal phase and the solution in \eqref{eq:mean_field_solution} is stable for the symmetry-broken phase
\begin{align} 
\label{eq:mean_field_solution}
A_\pm &= \kappa^2 \pm 4h^2, \quad B = \sqrt{\gamma_x^2 - \kappa^2} \,, \\
X &= \pm \frac1{\sqrt2 \kappa} \sqrt{A_- + \frac1{\gamma_x}B A_+} \,, \notag\\
Z &= \frac{2h}{\kappa^2}(\gamma_x - B) \,, \notag \\
Y &= \frac\kappa{2h} XZ \,. \notag
\end{align}
The point where the stable solution changes denotes the point of the phase transition,
it is given by Eq. \eqref{eq:qpt_position} in the article.

\end{document}